\documentclass{elsart}
\usepackage[square,comma]{natbib}
\usepackage{amsmath}
\usepackage{graphicx}
\usepackage{amsfonts}
\usepackage{amssymb}
\begin{document}
\runauthor{W. Rolke} \begin{frontmatter}
\title
{Correcting the Minimization Bias in Searches for Small Signals\thanksref{X}}
\author[wolf]{Wolfgang A. Rolke}
\address
[wolf]{Department of Mathematics, University of Puerto Rico - Mayag\"{u}%
ez, Mayag\"{u}ez, PR 00681, USA,
\newline Postal Address: PO Box 5959, Mayag\"{u}ez, PR 00681,
\newline Email: w\_rolke@rumac.uprm.edu}
\author[ang]{Angel M. L\'{o}pez}
\address[ang]{Department of Physics, University of Puerto Rico - Mayag\"{u}%
ez, Mayag\"{u}ez, PR 00681, USA}
\thanks
[X]{This work was partially supported by the Division of High Energy Physics of the US Department of Energy.}
\begin{abstract}
We discuss a method for correcting the bias in the limits for small signals if those limits were found based on cuts that
were chosen by minimizing a criterion such as sensitivity. Such a bias is commonly present when a "minimization" and an "evaluation" are
done at the same time. We propose to use a variant of the bootstrap to adjust the limits. A Monte Carlo study shows
that these new limits have correct coverage.
\end{abstract}
\end{frontmatter}
\newpage

\section{Introduction}

The standard method for quoting limits of small signals involves finding cuts
that eliminate as much background as possible while not reducing the number of
signal events significantly. A common measure for the performance of a certain
cut set is the sensitivity as discussed in Feldman and Cousins
\cite{Cousins-Feldman} and in Review of Particle Physics \cite{Particle Data
Group}. The sensitivity is defined as the average of the upper limits for an
ensemble of experiments all with the same cut set and with no true signal. We
adjust this concept by dividing by the detection efficiency and use this
''experimental sensitivity'' as our figure of merit. It can be thought of as a
measure for the size of an effect that could be discovered by a certain
experiment using a certain cut set. The smaller the sensitivity of a cut set,
the more likely we are to discover a signal that is truly present.

In recent years researchers have begun to realize that the choice of the
optimum cut set should be done ''blind'', that is, it should not be based on
the events in the signal region as this tends to bias the results by
overemphasizing statistical fluctuations. Instead one obtains an estimate of
the level of background in the signal region by either using real events not
in the signal region but assumed to have similar characteristics to those in
the signal region or simulated background events. We will call this sample of
events the background estimator sample.

Unfortunately, even in a blind analysis one is not safe from introducing
biases. The one we will consider in this paper stems from the fact that the
background estimator sample usually has limited statistics and there exists
the possibility of grossly underestimating the background by optimizing on a
negative statistical fluctuation. To be more precise, in the first analysis
step, which we shall call the minimization step, we choose a cut set by
searching through a large number of them and finding the one with the lowest
sensitivity. This procedure favors cut sets with low background levels. Having
settled on just one cut set we look into the signal region and see how many
events are still left after it is applied. Using this number and the estimated
background level, the signal limits are found. We shall call this second step
the evaluation step. The problem here is that the estimated background level
used in the evaluation step is that which resulted from the minimization step
which is systematically an underestimate.

This type of bias introduced by combining a minimization and an evaluation
step into one procedure is quite common in Statistics. As one example,
consider the problem of fitting a parametric curve to a histogram. Here we
usually start by estimating the parameters of the parametric function to be
fit, for example by finding the estimates of the parameters that yield the
lowest $\chi^{2}$. This is the minimization step. Then we want to know whether
our fit is sufficiently good, so we proceed to find the confidence level of
the $\chi^{2}$ statistic. This is the evaluation step. But in fact the
$\chi^{2}$, and therefore the confidence level, will be biased because the
parameter estimates were chosen to make the $\chi^{2}$ as small as possible.
In the next section we will give the results of a Monte Carlo study that shows
the presence of this type of bias in the search for small signals.

Of course we have known for almost a century how to adjust for this bias in
the case of the $\chi^{2}$, namely by adjusting the degrees of freedom of the
$\chi^{2}$ distribution. Unfortunately, in general it is very difficult to
find an analytic correction for this type of bias. In this paper we will
propose a numerical method, namely a variant of the bootstrap, for this
purpose. The bootstrap was popularized about 15 years ago, and it is fair to
say that it has revolutionized Statistics during the last decade. We will give
a brief introduction to the bootstrap in this paper, and then we will show how
a variation of the bootstrap which we will call the dual bootstrap can be used
to adjust for the bias discussed above. Finally, a Monte Carlo is used to
study the performance of the dual bootstrap.

\section{The Bias}

The bias described in the introduction will always be present when a
minimization and an evaluation step are combined, but whether this bias is
large or small compared to the statistical error is a different question. In
the analysis of small signals the end result is typically a confidence limit,
either a two sided confidence interval or just an upper limit. Whether or not
a method to compute confidence intervals works correctly has to be judged
solely based on the true coverage rate of the limits. For example, if we
generate $1000$ data sets from a Poisson distribution with mean $\lambda$ and
then use some method to compute $90\%$ confidence intervals for $\lambda$ for
each of these $1000$ data sets, then at least $900$ of these confidence
intervals should contain the true value $\lambda$. If two or more methods with
correct coverage are available, then one may use other criteria to make the
choice of method. For example, in physics one might prefer to use a method
that never yields an empty interval, or one might prefer a method that yields
on average, the shortest intervals. Such a choice has to made before examining
the data, of course.

For discrete distributions (like the Poisson) we have the added difficulty
that we can not achieve the desired coverage rate precisely. We will instead
require that the true coverage rate be at least as large as the nominal rate
for any value of the parameter, even if that will result in overcoverage for
many of the values.

We will use the method of Rolke and L\'{o}pez \cite{Rolke-Lopez} to compute
the confidence intervals. This is the only published method that treats the
uncertainty in the background rate as a statistical error. Like Feldman and
Cousins \cite{Cousins-Feldman} it solves the ''flip-flop'' problem, and it
always results in physically meaningful limits. Later Feldman and Cousins
\cite{Feldman2} independently solved the problem of an unknown background rate
by proposing a modification to the unified method. The bias problem described
here as well as its solution, though, do not depend on what method of
computation is used for either the sensitivity or the limits. As long as there
is some uncertainty in the background rate the bias would be equally present
if we had used for example Feldman and Cousins \cite{Cousins-Feldman} with
their modification or a Bayesian method.

To get an idea of the size of the bias we performed a Monte Carlo study of the
analysis of the $D^{0}\rightarrow\mu^{+}\mu^{-}$ decay using the data from
FOCUS \cite{link}. One problem in doing this MC is obtaining a large sample of
background events. In our study this sample was obtained by assuming the
background was due to other particles being misidentified as muons. A large
sample resulted from applying all base cuts except for muon identification to
a set of real data events. In this sample the events were weighted by the muon
misidentification probabilities which had been determined independently.
Although this may not be a perfect characterization of the real background, it
is good enough to study the question at hand which is more related to the
statistics than to the physics. The dimuon signal was generated with the full
FOCUS simulation program.

Fake data sets were generated by randomly choosing $M$\ events from the
simulated signal set and $N$ events from the background set. For the purpose
of this study we chose the number of background events $N$ from a Poisson
distribution with mean $16$ as in the real dimuon data, and $M$ was chosen
from a Poisson distribution with mean $\lambda$, where $\lambda$ was varied
from $0$ (meaning no signal was present) to $15$. For each value of $\lambda$
we generated $500$ fake data sets in this manner. To each of these data sets
we applied each of $13122$ cuts. The cuts used for this simulation were the
same cuts that had previously been chosen as appropriate for this analysis,
that is, each individual cut took on a reasonable range of values. In the next
step we found the cut that had the lowest sensitivity. This cut was then
applied to the signal region and the Rolke-L\'{o}pez method was used to find
the corresponding confidence limits. Finally those confidence limits were used
to calculate the true coverage rates. To make sure that any observed bias is
really due to the minimization-estimation problem, we also randomly chose $9$
individual cuts and always applied those same cuts to the fake data. In this
case no minimization was done, and hence no bias was expected.

The results of this MC study are shown in figure 1. As expected the limits for
the individual cuts have correct coverage, with the true coverage not dropping
much below the nominal value of $0.9$. That a few of the coverage rates on the
right side of the graph are below the 0.9 line is due to random fluctuations
in the MC as well as the discrete nature of the Poisson distribution. The
apparent drop in the coverage rates from the left to the right does not
continue, with the rates for $\lambda^{\prime}s$ larger than $6$ all just
above $0.9$. This was verified by running the MC for various values of
$\lambda$ up to $\lambda=15$.

Correct coverage is not the only characteristic an optimum methodology should
have. It is also important to obtain the strictest limits possible. That is
what the minimum sensitivity cut methodology attempts to do but in the limits
for these cuts we clearly see coverage rates well below the nominal rate. The
graph is based on 15 values for $\lambda,$ and it would be pure coincidence if
the lowest true coverage were obtained for one of those values. Therefore the
worst coverage should be expected to be well below the worst one observed of
about $0.845$. We can therefore conclude that we have a sizable bias in our
confidence limits due to the minimization-evaluation problem.

\section{A Brief Introduction to the Bootstrap}

The bootstrap method is a non-parametric alternative for finding error and
bias estimates in situations where the assumption of a Gaussian distribution
is not satisfied and where it is difficult or even impossible to develop an
analytic solution. Note that we are referring to the statistical method and
not to the S-matrix bootstrap as used in quantum field theory. In this section
we will present the reasoning behind the statistical bootstrap method and how
it is applied in practice.

Let us assume we are interested in estimating a certain parameter $\theta$
such as the width of a signal or a branching ratio. Let us also assume that we
have observations $X_{1},..,X_{n}$ from a distribution $F$ that depends on
$\theta$. Furthermore we have a method for finding an estimate $\widehat
{\theta}$ of $\theta$, say $\widehat{\theta}=T(X_{1},..,X_{n})$. The estimator
$T$ might be as simple as computing the mean of the observations or as
complicated as fitting a Dalitz plot.

Now, in addition to $\widehat{\theta}$ we will also need an error estimate as
well as an idea of the bias in the estimator $T$. If $T$ is fairly simple we
might be able to find its distribution and get an error and a bias estimate
analytically. If the situation is more complicated we might instead try a
Monte Carlo study. To do this we would simulate sampling from the distribution
$F$, generating many (say $k$) independent samples of size $n$, apply the
estimator $T$ to each and thereby get a sample of estimators $\widehat{\theta
}_{1},..,\widehat{\theta}_{k}$. Then we can look at a histogram of the
estimators, compute their standard deviation, and so on.

But what can we do if we do not know the distribution $F$? In that case the
data $X_{1},..,X_{n}$ is all we have, and any analysis has to be based on
these observations. The best estimate of the distribution function $F(x)$ is
the empirical distribution function $\widehat{F}(x)$ given by $\widehat
{F}(x)=\frac{1}{n}\cdot\left(  \text{number of observations }X_{i}\leq
x\right)  $, \ \ $x\in\Re,$ that is the percentage of events smaller than $x$.
Figure 2 shows a Gaussian distribution together with the empirical
distribution function of a sample of size $25$. The basic idea of the
bootstrap is to replace the distribution $F$ in the MC study above by its
empirical distribution function $\widehat{F}$. How does one sample from the
empirical distribution function? It is easy to show that this amounts to
sampling \textit{with replacement} from the observations $X_{1},..,X_{n}$. So,
if the (ordered) observations are $2.3,2.7,3.9,4.8,5.6,6.4$, then a bootstrap
sample might be $2.3,2.3,4.8,5.6,5.6,5.6$. A bootstrap sample has the same
sample size as the original data. It might include an original observation
more than once (such as $5.6$ in our example) or not at all (such as $3.9$).
As in the MC study, we will draw many (say $B$) of these bootstrap samples,
apply the estimator $T$ to each of them and thereby get bootstrap estimates
$\widehat{\theta}_{1}^{\ast},..,\widehat{\theta}_{B}^{\ast}$ of $\theta$. We
can then study these bootstrap estimates to get an idea of the error and the
bias of $T$.

The bootstrap method as described above was first developed by \ B. Efron in
\cite{Efron}. Since then a great deal of theoretical work has been done to
show why and when the bootstrap method works, see for example Hall
\cite{Hall}, and it has been successfully used in a wide variety of areas.
Previous applications of the bootstrap in High Energy Physics can be found in
Hayes, Perl and Efron \cite{Hayes}\ and in Alfieri et al. \cite{Alfieri}. For
a very readable introduction to the subject see Efron and Tibshirani
\cite{Efron and Tibshirani}.

\section{The Dual Bootstrap and Bias Corrected Limits}

The bias observed in section 2 comes from combining a minimization and an
evaluation step. We will use a version of the bootstrap to decouple those two
steps. One possible solution for this bias problem would be to use a split
sample approach: Randomly divide the data set into two pieces, use one part to
find the cut with the smallest sensitivity and use the other part to find the
limits of the branching ratio. Unfortunately this method has a number of
problems: first there is the question how large a portion of the sample should
be used to find the best cut. There does not appear to be any theory guiding
this choice at this time. Another difficulty in searches for small signals is
that we are already working with very few events, and to split those up even
further does not seem to be a good idea. Finally the split sample method
introduces an additional random error: because we only have very few events in
the sideband it might well make a difference whether a specific event ends up
in the sample used to find the best cut combination, or whether it gets put
into the sample used to find the limits.

All these problems can be avoided if we proceed as follows: we draw one
bootstrap sample from the data and find the cut with the smallest sensitivity
for this bootstrap sample, then we will draw another bootstrap sample,
conditional on the data but independent from the first, to find the limits.
This procedure will then be repeated $B$ times, with a $B$ of about $5000$. In
this manner we will get $B$ lower and upper limits. Finally we will use the
median of the lower and the median of the upper limits as our estimates. We
use the median because it is less sensitive than the mean to a few unusually
large observations. Also, in the case where the signal rate is zero, if even a
few of the $B$ bootstrap estimates of the lower limit are positive, the mean
would also be positive (and wrong), whereas the median is still zero (and
therefore correct).

In this way for each bootstrap sample we get a cut set that is optimal for the
first bootstrap sample but not necessarily for the second, which is
representative of the underlying distribution. We can therefore expect to get
unbiased estimates of the limits or, in other words, limits with the correct
coverage rate.

We repeated the MC study discussed in section 2, now using the dual bootstrap
method. Figure 3 shows that the dual bootstrap method yields limits with the
correct coverage, effectively removing the minimization-evaluation bias.
Similar MC studies with different nominal coverage rates and different
background rates, both smaller and larger than the rate of $16$ shown here,
have confirmed this conclusion.

\section{Presenting Results}

We would recommend to publish two graphs to present the results of a dual
bootstrap analysis. On the one hand we suggest a scatter plot of the upper
confidence limits versus their experimental sensitivities for each cut
combination, together with lines showing the medians of the dual bootstrap
upper limits and sensitivities. This graph should show that the quoted results
are ''reasonable'' when compared to what would have been quoted if only one
cut combination had been used, that is, the lines should be inside the range
of values from the individual cuts. On the other hand we would suggest to plot
the histograms of the bootstrap estimates of the lower and upper limits, so
that the reader can get an idea of the variation in those limits. As an
illustration we have generated two fake data sets, similar to those used in
the simulation study above. In the first case we generated a data set without
a signal and then ran the dual bootstrap $5000$ times. The median bootstrap
lower limit is $0$, and so we draw the histogram of the bootstrap upper limits
only, together with the median bootstrap upper limit and the median bootstrap
sensitivity. The result is in figure 4. In figure 5 we have the results for a
data set with a signal rate of $6.0$. Here we draw both the histograms of the
lower and the upper bootstrap limits, together with the corresponding medians.

Another advantage of the bootstrap method is apparent from these graphs,
namely that we can get an idea of the variation in the confidence limits and
the sensitivities. That confidence limits are themselves random entities and
therefore have an error is often underappreciated. These errors can be easily
computed from the bootstrap samples, although it would be advisable to use a
robust estimate of the standard deviation rather than the usual formula,
because the bootstrap estimates are clearly skewed to the right. A good
estimator of the error would be the interquartile range, defined by $(75^{th}%
$percentile - $25^{th}$percentile)/$1.35$. The constant $1.35$ here makes the
interquartile range equivalent to the standard deviation in the case of
Gaussian data. These errors might be useful in comparing and combining the
confidence intervals of different experiments.

\section{Claiming a Discovery}

An important issue is the question on how to decide whether one should claim a
discovery. From a statistical point of view this should be done by performing
a hypothesis test with the null hypothesis $H_{o}$:''\textit{There is no
signal present}''. Because of the duality of confidence intervals and
hypothesis tests we can actually do this by simply finding confidence
intervals with the appropriate $\alpha$, say $\alpha=99.73\%$. corresponding
to a $3\sigma$ effect or $\alpha=99.99994\%$. corresponding to a $5\sigma$
effect, and deciding that a signal is real if the lower limit is greater than
$0$. By running the dual bootstrap with a variety of $\alpha$'s we can even
get an estimate of the size of the effect, or what is called a p-value in
statistics, namely the probability of claiming a signal when there really is
none. As an example consider one the data sets used\ in our simulation study.
This data set was generated with a signal rate of $6.0$. The dual bootstrap
estimate of the lower limit was $0$ for $\alpha=0.99991$ and $0.031$ for
$\alpha=0.99992$, based again on $5000$ bootstrap runs. By equivalence to
confidence intervals from a Gaussian distribution this amounts to a
$3.93\sigma$ effect.

If a signal is present, an obvious estimator for the signal rate is the
maximum likelihood estimator $\widehat{\lambda}=\max(x-y/\tau,0)/\varepsilon$,
where $x$ is the number of events in the signal region, $y$ is the number of
background events, $\tau$ is the size of the background region relative to the
size of the signal region and $\varepsilon$ is the detection efficiency of the
cut. Of course this estimate suffers from the same cut-selection bias, and
would generally lead to estimates of the signal rate that are too large. Again
we can use the dual bootstrap method to adjust for this bias, by computing
$\widehat{\lambda}$ for each dual bootstrap sample and using the median as the
point estimate.

\section{Conclusion}

We have shown that selecting a cut set based on the smallest sensitivity and
then using it to find the limits of the branching ratio (or generally of any
other quantity) introduces a bias. In the case of the $D^{0}\rightarrow\mu
^{+}\mu^{-}$ decay in FOCUS a MC study indicates that this bias is quite
large, certainly too large to ignore. It is reasonable to assume that a
sizable bias is present whenever the minimization criterion is related to the
evaluation criterion, for example if both are related to the signal to noise
ratio. We have developed a method based on the bootstrap technique from
Statistics that corrects for this type of bias. A MC study for the
$D^{0}\rightarrow\mu^{+}\mu^{-}$ decay shows that this new method performs
very well. Showing the histograms of the bootstrap samples together with the
median also gives an idea of the variation in these estimates.

FORTRAN routines for the dual bootstrap method as well as for computing the
Rolke-L\'{o}pez limits are available from the authors by sending an email to w\_rolke@rumac.uprm.edu.

\section{Acknowledgements}

We would like to thank Daniel Engh of Vanderbilt University for suggesting the
scatter plot which is very useful for presenting the results.

\section{Appendix}%

%TCIMACRO{\FRAME{ftbpFU}{403pt}{316.375pt}{0pt}{\Qcb{True coverage rates for
%individual cuts (dotted lines) and for minimum sensitivity cut (solid line).
%The rates for the individual cuts are always larger than the nominal rate as
%is desired, wheras the true coverage for the minimum sensitivity is often well
%below the nominal rate of 0.9.}}{}{figure1.eps}%
%maintain-aspect-ratio TRUE;  display "USEDEF";  valid_file "F";  width 403pt;
%height 316.375pt;  depth 0pt;  original-width 7.7262in;
%original-height 5.9551in;  cropleft "0";  croptop "1.1607";
%cropright "1.1408";  cropbottom "0";
%filename 'figure1.eps';file-properties "XNPEU";}}}%
%BeginExpansion
\begin{figure}
[ptb]
\begin{center}
\includegraphics[
trim=0.000000in 0.000000in -1.087849in -0.956984in,
height=316.375pt,
width=403pt
]%
{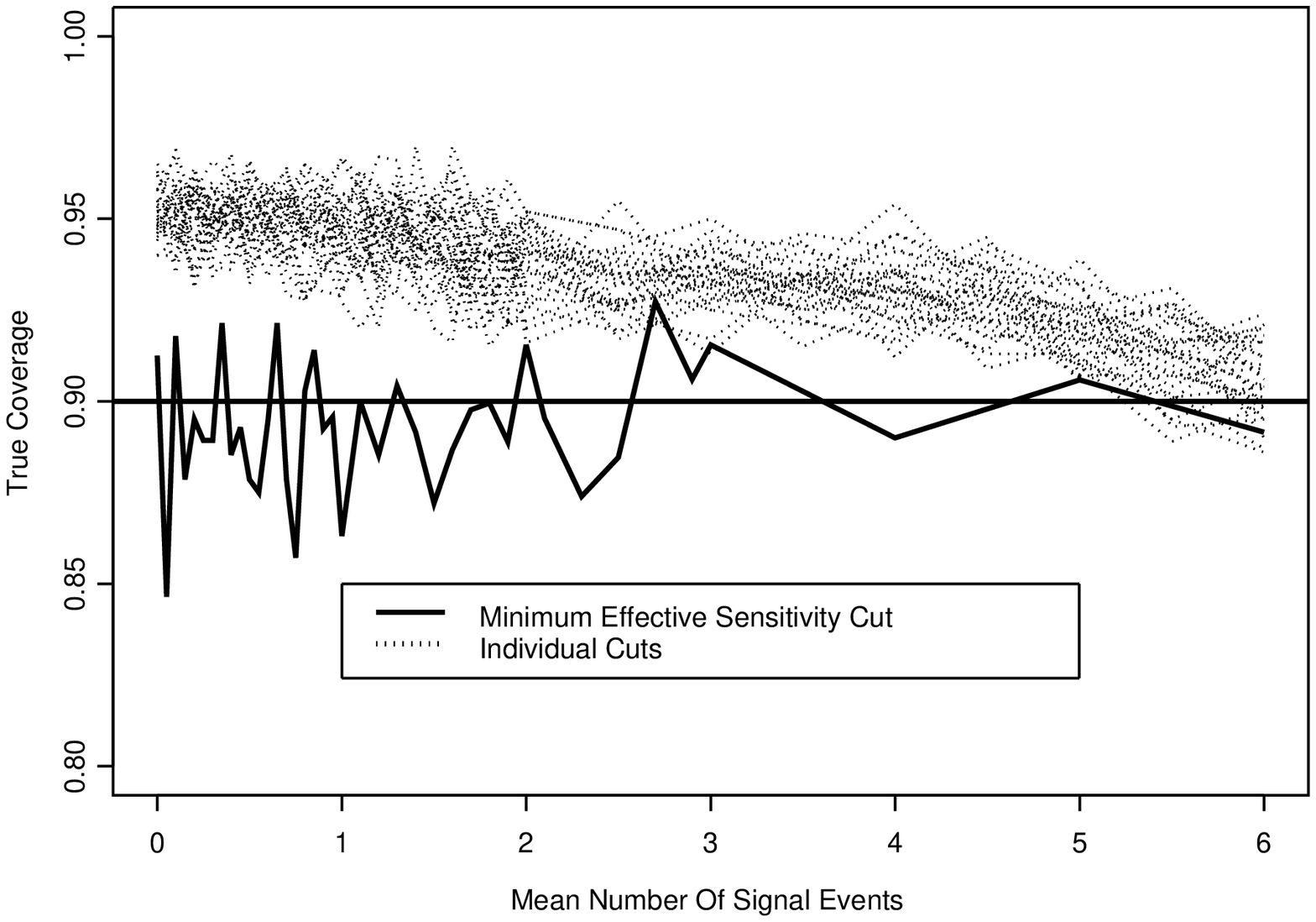}%
\caption{True coverage rates for individual cuts (dotted lines) and for
minimum sensitivity cut (solid line). The rates for the individual cuts are
always larger than the nominal rate as is desired, wheras the true coverage
for the minimum sensitivity is often well below the nominal rate of 0.9.}%
\end{center}
\end{figure}
%EndExpansion%

%TCIMACRO{\FRAME{ftbpFU}{403pt}{316.375pt}{0pt}{\Qcb{Exact distribution
%function (smooth function) and empirical distribution function (step function)
%of a sample of size 25 from a Gaussian distribution.}}{}{figure2.eps}%
%maintain-aspect-ratio TRUE;  display "USEDEF";  valid_file "F";  width 403pt;
%height 316.375pt;  depth 0pt;  original-width 7.7262in;
%original-height 5.9551in;  cropleft "0";  croptop "1.1389";
%cropright "1.1192";  cropbottom "0";
%filename 'figure2.eps';file-properties "XNPEU";}}}%
%BeginExpansion
\begin{figure}
[ptb]
\begin{center}
\includegraphics[
trim=0.000000in 0.000000in -0.920963in -0.827164in,
height=316.375pt,
width=403pt
]%
{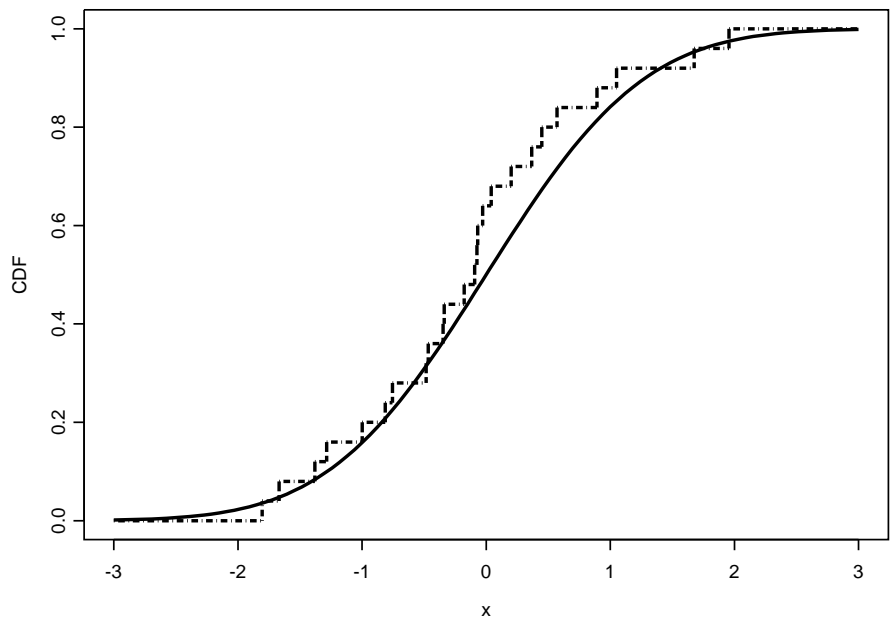}%
\caption{Exact distribution function (smooth function) and empirical
distribution function (step function) of a sample of size 25 from a Gaussian
distribution.}%
\end{center}
\end{figure}
%EndExpansion%

%TCIMACRO{\FRAME{ftbpFU}{403pt}{316.375pt}{0pt}{\Qcb{True coverage rates for
%the limits of the signal rate based on the cut set with the smallest
%sensitivity (solid line) and for the dual bootstrap method. Clearly the dual
%bootstrap correctly adjusts the limits and yields correct coverage.}}%
%{}{figure3.eps}{\special{ language "Scientific Word";  type "GRAPHIC";
%maintain-aspect-ratio TRUE;  display "USEDEF";  valid_file "F";  width 403pt;
%height 316.375pt;  depth 0pt;  original-width 7.7262in;
%original-height 5.9551in;  cropleft "0";  croptop "1.1534";
%cropright "1.1335";  cropbottom "0";
%filename 'figure3.eps';file-properties "XNPEU";}}}%
%BeginExpansion
\begin{figure}
[ptb]
\begin{center}
\includegraphics[
trim=0.000000in 0.000000in -1.031448in -0.913512in,
height=316.375pt,
width=403pt
]%
{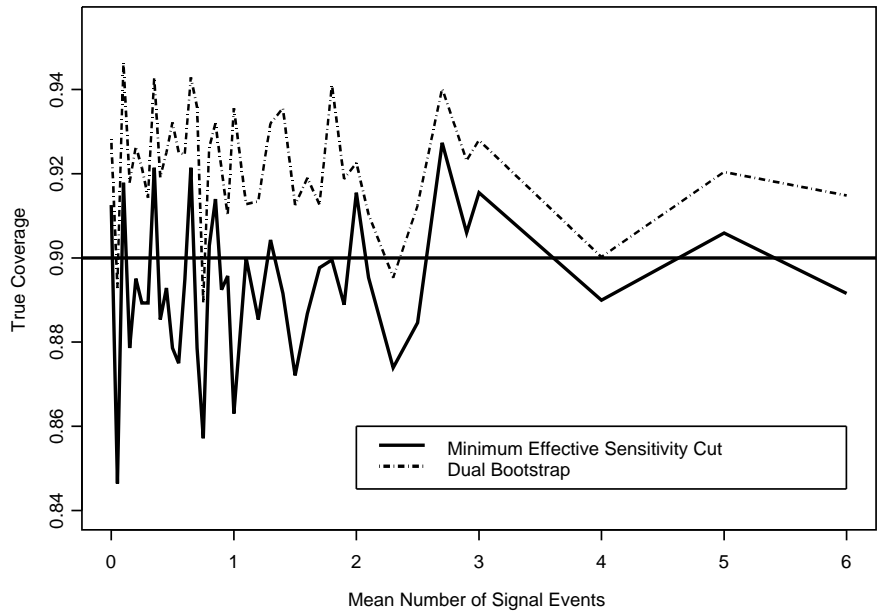}%
\caption{True coverage rates for the limits of the signal rate based on the
cut set with the smallest sensitivity (solid line) and for the dual bootstrap
method. Clearly the dual bootstrap correctly adjusts the limits and yields
correct coverage.}%
\end{center}
\end{figure}
%EndExpansion%

%TCIMACRO{\FRAME{ftbpFU}{403pt}{316.375pt}{0pt}{\Qcb{Left Panel: Scatter plot
%of the $90\%$ upper limit for the number of signal events versus the
%experimental sensitivity for all of the cut combinations. The solid lines
%correspond to the median dual bootstrap upper limit (horizontal) and
%sensitivity (vertical). Right Panel: Histogram of the upper limit for $5000$
%dual bootstrap samples. Both the median upper limit and sensitivity are marked
%with vertical lines.}}{}{figure4.eps}{\special{ language "Scientific Word";
%type "GRAPHIC";  maintain-aspect-ratio TRUE;  display "USEDEF";
%valid_file "F";  width 403pt;  height 316.375pt;  depth 0pt;
%original-width 7.7262in;  original-height 5.9551in;  cropleft "0";
%croptop "1.0110";  cropright "0.9938";  cropbottom "0";
%filename 'figure4.eps';file-properties "XNPEU";}}}%
%BeginExpansion
\begin{figure}
[ptb]
\begin{center}
\includegraphics[
trim=0.000000in 0.000000in 0.047903in -0.065506in,
height=316.375pt,
width=403pt
]%
{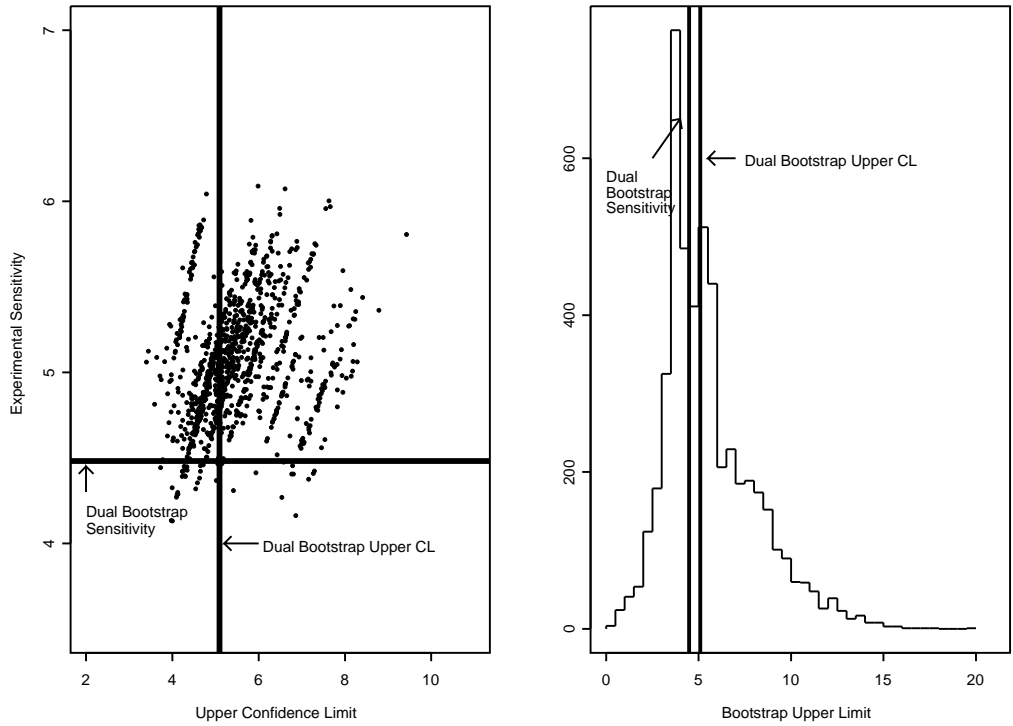}%
\caption{Left Panel: Scatter plot of the $90\%$ upper limit for the number of
signal events versus the experimental sensitivity for all of the cut
combinations. The solid lines correspond to the median dual bootstrap upper
limit (horizontal) and sensitivity (vertical). Right Panel: Histogram of the
upper limit for $5000$ dual bootstrap samples. Both the median upper limit and
sensitivity are marked with vertical lines.}%
\end{center}
\end{figure}
%EndExpansion%
%TCIMACRO{\FRAME{ftbpFU}{403pt}{316.375pt}{0pt}{\Qcb{Left Panel: Scatter plot
%of the $90\%$ upper limit for the number of signal events versus the
%experimental sensitivity for all of the cut combinations. The solid lines
%correspond to the median dual bootstrap upper limit (horizontal) and
%sensitivity (vertical). Right Panel: Histograms of the lower and upper limits
%for $5000$ dual bootstrap samples. Both the median lower and upper limits are
%marked with vertical lines.}}{}{figure5.eps}%
%maintain-aspect-ratio TRUE;  display "USEDEF";  valid_file "F";  width 403pt;
%height 316.375pt;  depth 0pt;  original-width 7.7262in;
%original-height 5.9551in;  cropleft "0";  croptop "1.0319";
%cropright "1.0145";  cropbottom "0";
%filename 'figure5.eps';file-properties "XNPEU";}}}%
%BeginExpansion
\begin{figure}
[ptbptb]
\begin{center}
\includegraphics[
trim=0.000000in 0.000000in -0.112030in -0.189968in,
height=316.375pt,
width=403pt
]%
{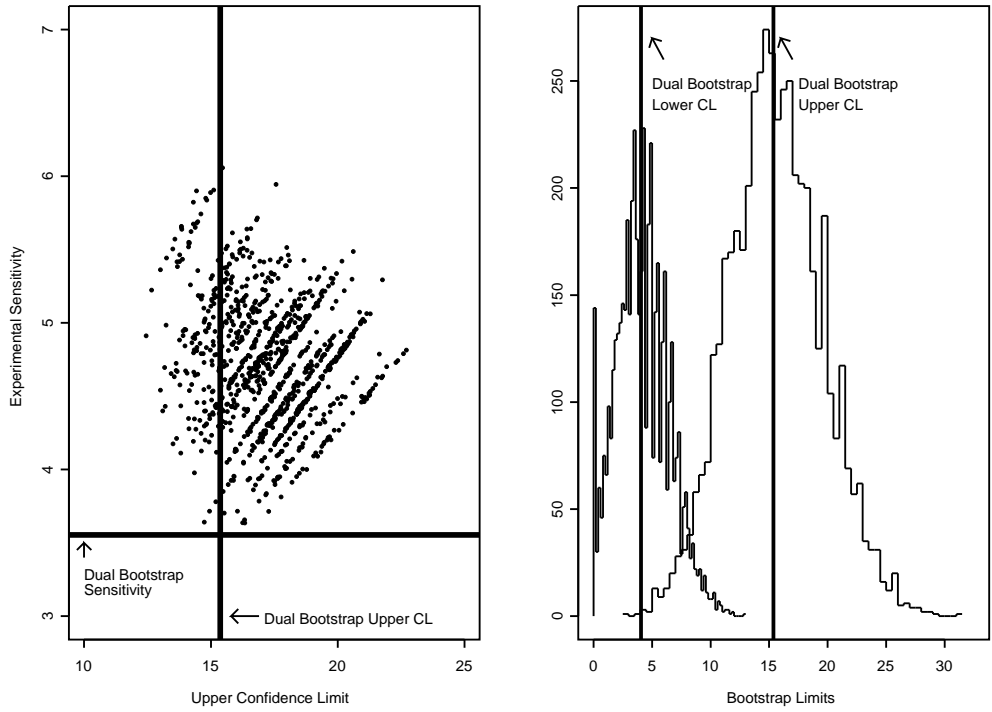}%
\caption{Left Panel: Scatter plot of the $90\%$ upper limit for the number of
signal events versus the experimental sensitivity for all of the cut
combinations. The solid lines correspond to the median dual bootstrap upper
limit (horizontal) and sensitivity (vertical). Right Panel: Histograms of the
lower and upper limits for $5000$ dual bootstrap samples. Both the median
lower and upper limits are marked with vertical lines.}%
\end{center}
\end{figure}
%EndExpansion
\end{document}